# Sprouting technology otherwise, *hospicing* negative commons

Rethinking technology in the transition to sustainability-oriented futures

Martin Deron
Concordia University
Montréal, Québec, Canada
martin.deron@mail.concordia.ca

## ABSTRACT

Due to its significant and growing environmental harms, both directly through its materiality and indirectly through its pervasive integration into unsustainable economic systems, ICT will need to be radically redirected to align with sustainability-oriented futures. While the role of ICT in such futures will likely diverge significantly from current dynamics, it will probably not be entirely disconnected from the present. Instead, such transition involves complex dynamics of continuity, adaptation and rupture. Drawing from recent work in transition studies, the commons (particularly "negative commons"), as well as some of the *Limits* literature, this article proposes a conceptual framework for navigating this redirection. The framework attempts to bring together the disentanglement from sociotechnical elements incompatible with long-term sustainability and the support of existing practices that may serve as foundations for alternative technological paths. It introduces four categories: *ruins, ghosts, seeds* and *visions*, to examine how material and cultural aspects of computing may become obsolete, persist in latent or reinterpreted forms, or contribute to sustainability-oriented futures. Through both empirical and speculative examples, I intend to show how this lens can help researchers and practitioners engage more concretely with the tensions, inheritances, and opportunities involved in redirecting computing towards more sustainable and equitable futures.

## KEYWORDS

Negative commons, sustainability transition, technology otherwise, transition studies

## 1  INTRODUCTION

Far from the dematerialized claims that often accompany it, the expanding digitalization of societies is grounded in a growing "Technosphere" (Zalasiewicz et al., 2017), composed of over 30 billion connected devices, 80 million servers, hundreds of thousands of kilometers of fiber-optic cables and vast quantities of networking equipment worldwide (Aubet et al., 2025). This gigantic infrastructure shows the profoundly material nature of the so-called digital world. The manufacturing, use and disposal of all of this infrastructure reveal environmental consequences frequently overlooked in computing research, including in sustainability-oriented studies (Ligozat et al., 2022). They exert considerable pressure on natural resources and the stability of biophysical systems, contributing to rising greenhouse gas (GHG) emissions (estimated at around 3-4% of global emissions [Freitag et al., 2021; Aubet et al., 2025]), increasing energy demands (Gelenbe, 2023), and significant water consumption (Valdivia, 2024; Gonzalez Monserrate, 2024).

Moreover, as "enabling technologies" (Hilty & Aebischer, 2015), Information and Communication Technology (ICT) also contributes to amplifying unsustainable societal dynamics, operating within growth-oriented market dynamics. While certain cases have the potential to improve energy or resource efficiency locally (Rolnick et al., 2022), these gains are typically reinvested to expand production and profits, thereby exacerbating environmental pressure more broadly.

Digital technologies thus become embedded in economic processes that are at the root of ecological degradation (Longaretti & Berthoud, 2021; Martineau and Durand Folco, 2023). Comber and Eriksson (2023) deepen this critique by framing computing as a potential case of "ecocide", emphasizing its role not only in direct environmental harms but more pervasively as an enabler of further harms through its integration into the most environmentally-damaging sectors. The authors argue that computing's most prevalent, widespread and severe harms come from its role as an accelerator of destructive practices (p.6).

Building on this emerging literature, there is a growing realization, both within computing research and beyond, that ICT, like any other sector, is tied with ecological limits. Its very existence relies on the continuous extraction of metals and minerals and requires significant water and energy resources. On the other hand, ICT infrastructure is particularly vulnerable to the impacts of extreme climate events, such as floods and droughts. This dual role, both contributing to and suffering from ecological degradation, has led to calls for a radical reorientation of computing research in line with long-term sustainability (Nardi et al., 2018)

In response, emerging "technologies otherwise"[1] communities and networks offer glimpses of alternative technological paths. These include efforts to detach innovation from growth (Pansera & Fressoli, 2021), explore what constitutes a "convivial technology" (Vetter, 2018), and a solar-powered website that challenges the need for availability (Roscam Abbing, 2021). These projects point to the possibility of radically different technological imaginaries and materializations.

---

[1] See for example the "Tech Otherwise – Another Tech is Possible" collective at https://techotherwise.pubpub.org/, active since 2019

While thinking about different technological futures is essential for pluralizing narratives so often dominated by hegemonic views (Bugeau and Ligozat, 2022), it seems also important to bridge these alternative imaginaries with present practices. Although the future remains uncertain, it is unlikely to be entirely disconnected from the present. Transition studies show that systems change typically follows a gradual path of destabilization and reconfiguration (Geels, 2019). As broader socio-economic contexts change, once-marginal alternatives can gain visibility, support and become more dominant.

From this perspective, imagining future societies organized around ecological and social viability (and not economic growth) requires tackling continuity and rupture together. ICT systems incompatible with emerging socio-ecological realities may evolve, adapt, or disappear, all the while smaller-scale or peripheral systems aligned with the new paradigm may become more prominent.

This article seeks to advance the discussion on redirecting ICT toward sustainability-oriented futures in a way that fosters the creation of new possibilities together with the difficult task of untangling currently dominant practices as they become increasingly inadequate. Drawing inspiration from transition studies and recent work on the commons, especially anti-capitalist and negative commons, I propose a new conceptual framework for assessing the directionality of technological systems based on their compatibility with such futures. After clarifying the main theoretical inspirations of this work, I identify four categories to classify technological legacies and emergences in the context of socio-ecological transitions: "ruins", "ghosts", "seeds", and "visions" and attempt to provide both existing and speculative examples of their manifestation. Lastly, I finish this article with different ways researchers and practitioners can use the proposed framework to help inform and potentially redirect their practice.

## 2 THEORETICAL INSPIRATIONS

### 2.1 Transition studies

The field of transition studies has produced different models to describe how societies change over time (both as a reflective practice and in order to inspire change) such as the three horizons model (Sharpe et al., 2016) and the multi-level perspective (MPM) framework (Geels, 2019). One such model is the Berkana Institute's *Two Loop Model* (2011), illustrated in Figure 1 below. It depicts how dominant, society-wide systems, gradually lose their hegemony and create space for the emergence of a new paradigm. The model illustrates how this shift involves different phases to accompany the ending and the death of the old system as well as the creation of the new one.

While the two loops (showing the declining system and the emergent one) do not intersect due to their grounding in distinct paradigms, the transition from one to the other is not framed as a total rupture or revolution. Rather, elements of the dominant system undergo transformation, adapting to the emerging paradigm, while others that no longer serve the new context follow a process of *hospicing* and ultimately *die* or are *composted* to be used under a new form in the emergent system. This model represents systems change under what is often described as a gradual "process of emergence" (Beckerman, 2022).

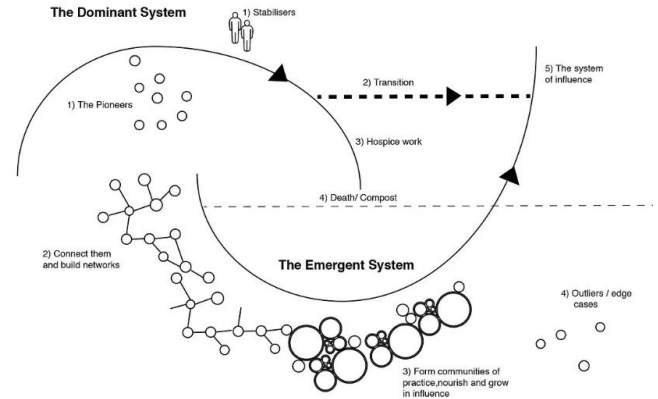

Figure 1 - Two Loop Model, Berkana Institute (2011), illustration by Robinson (2019)[2]

A critical contribution from the model lies in its articulation of the many different responsibilities required of change agents. While a lot of the work that is expected goes into inspirational and visionary dimensions of change (such as exploring and creating alternatives or connecting, nourishing communities of practice and building networks), there is also critical work to be done in stabilizing, accompanying and *composting* the unsustainable aspects of the dominant system. This work is vital to minimize the negative consequences of a system-wide decline and preparing ground for more sustainable structures to flourish. Lastly, there is also an important need for change agents to support transition from the previous model to the emergent one, building bridges and opportunities to cross over.

### 2.1 Commons beyond capitalism

Amidst growing interest in alternatives to capitalism, fueled by worsening socio-ecological crises, the concept of "commons" has gained renewed attention among scholars and activists. As enclosure mechanisms intensify under capitalism, there has been a resurgence of efforts to defend existing commons and establish new ones as acts of resistance (García-López et al., 2021).

In academia, the commons have been approached from various angles. While Garrett Hardin's *Tragedy of The Commons* (1968) framed the commons as inherently vulnerable to overuse, Elinor Ostrom's analysis (1990) offered a counter-narrative, documenting successful models of community self-governance.

A more critical tradition has also examined the commons through the lens of dispossession and enclosure, understood as the systematic removal of communal access by capitalist expansion, often through violence (Federici, 2004; Luxemburg, 2015; Marx, 2014). This critical approach positions the commons not only as resource management models, but as vehicles for transformative social relations. Caffentzis and Federici (2014) defend the importance of producing commons that are truly "anti-capitalist"; conceived as "both autonomous spaces from which to reclaim control over the conditions of our reproduction, and as bases from which to counter the processes of enclosure and increasingly disentangle our lives from the market and the state" (p. 101). In contrast to "co-opted" or service-oriented commons, seen by the authors as mere buffers against "the destructive nature of neo-liberalism" (p. 100), anti-capitalist commons aim to dismantle capitalist dynamics at their core in a way that makes a non-capitalist world even possible to exist (p.103).

---

[2] The Two Loop Model developed by the Berkana Institute was explained in a video that no longer seems accessible as a primary source. However, many have used the model to illustrate it graphically in very similar ways. This illustration was elaborated by Robinson, C. (2019) and can be found at https://medium.com/thefarewellfund/hospicing-the-old-16e537396c4b

## 2.2 Negative commons

While the idea of the commons has been influential in rethinking social relations and re-localizing production, it often remains tied to a development logic, assuming commons are inherently good and community-enhancing (Maurel, 2023). Yet, the realities of ecological breakdown force us to reckon with the limits of human activity, especially where it threatens planetary boundaries (Raworth, 2018; Bärnthaler & Goud, 2023).

Some infrastructures, activities, or belief systems, deeply embedded in our economics and cultures (such as always accessible "cloud" infrastructures) may no longer be viable for sustainable futures. This raises difficult questions: how do we deal with shared resources or legacies that are not beneficial, and in some cases actively harmful? Whether we like or not, these realities will require forms of management and care even as they lose relevance.

Here, the concept of "negative commons" offers a useful lens. Originally introduced by Mies and Bennholdt-Thomson (2001), the term was used to describe organic waste that once circulated locally but became unmanageable with the rise of mechanized, privatized agriculture, thereby disrupting communal and circular activities

Philosopher Alexandre Monnin expands on this idea to include the physical and epistemic legacies of capitalism: nuclear waste, polluted landscapes, concreted cities; all persistent remnants of economic activity that no traditional actors want to claim responsibility for (Monnin, 2023b, p.38). As Monnin explains it, these are not simply byproducts, they are "ruins": residues of outdated systems that resist decay and continue to impose costs and harms, long after they have stopped serving their original purpose. From the perspective of the Anthropocene, much of our material infrastructure, such as roads, power grids, data centers but also digital platforms or tools, can be seen as ruins in waiting.

When thinking about sustainability transition, these ruins are important to consider since material stocks, such as infrastructure or machinery, have a strong influence on social practices and can lock societies into unsustainable trajectories (Plank et al., 2021). As they become ruins, people lose access to the benefits such infrastructures once offered, while still having deal with the impacts of their lingering presence. Mines, such as the ones exploited to collect rare minerals to manufacture digital devices or integrated circuits, hold financial value only when they are operational. Once the mine site shuts down, it leaves behind displaced and potentially poisoned land, toxic tailings, and long-term ecological harm (Izoard, 2024). These burdens largely fall on local communities who have little capacity to handle it. Monnin calls these realities "ruined ruins": infrastructures that outlast their utility but not their consequences.

Monnin contrasts these with "ruinous ruins" to qualify not the mine itself, but the systems and ideologies that led to its excavation in the first place. "What is most ruinous today", he writes, "are not open-pit mines, but the processes commanding that we dig them" (2023b, p.39)[3]. These include growth and accumulation-oriented economic models and generate the conditions for future "ruination". These "ruinous ruins" remind us of the importance of working with ideological deconstruction in addition to physical dismantling.

Importantly, these ruinous ruins often manifest materially too. Technological "lock-ins" (Seto et al., 2016), such as IoT infrastructures or proprietary software ecosystems, demonstrate how infrastructures constrain futures by making certain paths difficult to exit. Engaging with these ruins requires both critical understanding of their trajectories, and the construction of alternative pathways.

This engagement is inherently political. It requires shifting viability criteria away from growth and short-term returns on investment toward long-term ecological and social sustainability. As Paprocki notes: "anticipatory ruination justifies destruction in the present in anticipation of future threats" (2022, cited in Monnin, 2023b, p.81). This is what Monnin, in his book, calls "politicizing renouncement": a profound transformation of the system involves confronting future ruins in the making, to retool the non-viability of negative commons (2023b, p.83).

The framework of negative commons helps make visible the often-neglected conflictual dimensions of transitions: the artefacts we inherit, the futures they enable or block, and the difficult closures we must imagine. Anticipating ruination brings future consequences into present debates, encouraging us to identify and navigate the realities that threaten long-term sustainability, but escape responsibility under current systems. As Monnin writes: "Capitalism did not wait for us to learn how to close. What's new here is closure for reasons of non-viability" (2023a, p. 242). Identifying and politicizing these negative commons seems crucial to help transition in a way that is the most fair, democratic and least brutal possible.

## 2.3 Collapse informatics

Lastly, this contribution resonates with reflections from "Collapse Informatics", a well-known literature in the Limits community. Originally introduced by Tomlinson et al. (2013), Collapse Informatics is defined as "the study, design, and development of sociotechnical systems in the abundant present for use in a future of scarcity" (p.1). Like transition studies' attention to "niches" (Geels, 2019), Collapse Informatics explores present-day practices that deviate from dominant paradigms and which may offer empirical access into potential alternative futures. This includes what Fry (2009) terms "redirective practices".

---
[3] The book was published in French and the quotations included were translated by me.

This article builds on the same underlying premise that future contexts may involve greater scarcity and that it is crucial to use anticipation to try and shape more equitable and sustainable futures. Similarly, it also offers to look at existing practices to envision what future dynamics may look like in the transition to sustainable futures.

# 3 HOSPICING NEGATIVE COMMONS, SPROUTING ALTERNATIVES

When we look at ICT as a system, there is a clear dominant paradigm characterized by continuous and unquestioned growth, and accompanied by dynamics of accumulation, privatization and commodification, which is increasingly misaligned with socio-environmental crises (Comber & Eriksson, 2023). Therefore, it is important to identify which aspects may already constitute, or may eventually become, negative commons. This work seems necessary for stabilizing the dominant system as it begins to decline, and for anticipating as best as we can what must be managed, repurposed, or ultimately given hospice on the path to more sustainable futures.

On the other hand, considering the current hegemony of the dominant paradigm, it is not yet possible to define clearly what the emergent paradigm looks like. Perhaps looking for one alternative paradigm is a mistake too. The homogenous nature of the dominant system, designed by and for the Global North, expresses a very narrow idea of what technology looks like. In contrast, a sustainability-oriented emergent paradigm may be best represented by a network of different smaller and heterogenous systems, better connected to local value systems and environmental realities (akin to reflections on the "pluriverse" (Escobar, 2018)). In this sense, by looking at the peripheries and the margins of the current system, where it has historically caused the most harm and where communities have long been experimenting with different practices and ways of living, we can already identify existing networks, communities and uses that pave the way for a world *afterwards*.

## 3.1 Conceptualizing ruins, ghosts, seeds and visions

To help conceptualize the transition from the current growth-oriented paradigm to an emergent, sustainability-driven paradigm, I propose to engage with anticipatory ruination in a speculative manner. Presuming that ICT will follow a fundamental system change in order to become compatible with ecological limits, it might be useful to investigate current and projected digital realities under four categories: ruins, ghosts, seeds and visions.

Ruins and ghosts are derived from Monnin's work on negative commons, which I separate into two categories following more classical typologies: one focused on material elements (ruins) and a second dealing with cultural elements (ghosts). Furthermore, given the temporal jump induced by anticipatory ruination, I also put aside the distinction between "ruinous" and "ruined" ruins, instead focusing on the nature of the elements rather than their activity level.

To use this framework in a way that not only sheds light on the decline of the dominant paradigm but also helps envision the emergent paradigm(s), I have created two other categories that contrast directly with the first two. I propose to look at seeds (material elements) and visions (cultural elements): existing manifestations of ICT that may take on a larger role in the transition to sustainability-oriented futures. Figure 2 (on page 7) attempts to visually bring together the categories in one place with some of the examples mentioned in the following categories.

### 3.1.1 Ruins

Borrowing the term from Monnin (2023b), "digital ruins" are understood here as the material elements of ICT systems – such as infrastructures, components, devices or exploitation sites - that are likely to persist as negative commons in sustainability-oriented futures[4]. This first category aims to capture the many physical realities that may remain long after the end of the socio-technical systems that created them in the first place. These digital ruins already exist among what Storm (2014) refers to as "landscape scars": abandoned excavation sites, industrial buildings, or obsolete network equipment, marked by histories of environmental degradation and economic exploitation (Francaviglia, 1991). These structures endure, even after their original function has ceased, leaving behind both a visual legacy of the industrial forces that produced them (Fancaviglia, 2016) and long-lasting ecological and social consequences for human and non-human communities. They constrain certain forms of action while enabling others, exerting pressure on the (socio-political) landscape.

As the effects of climate degradation increase in both frequency and intensity, many ICT infrastructures will be vulnerable to disruption and decay. Extreme weather events such as wildfires, droughts, floods, landslides and strong winds (all projected to intensify in coming decades) pose serious risks to the integrity of key infrastructures, notably telecommunication systems, electricity transmission and distribution networks, and railway corridors (France Stratégie, 2024). These deteriorating physical structures represent incoming digital ruins: material remnants that will need to be dismantled, stabilized, accompanied or *hospiced* – to use the language of the Two Loop model.

These evolving environmental realities are likely to spur not only infrastructural breakdown but also social and political reconfigurations, especially as perceptions of resource abundance begin to shift (admittedly a very Global North-centered perspective). In projected environmental futures of resource scarcity, infrastructures that were once seen as symbols of progress or economic opportunity, such as data centers, may become subject of unprecedented political contestation. Historically positioned outside the realm of public political debate, the maintenance or installation of data centers may become the object of intense public deliberation, raising questions of resource allocation and collective values. Why should a data center have privileged access to limited water and electricity supplies? What social function justifies such access? Who ultimately benefits from it?

---

[4] It is important to note here that Monnin strongly advocates against an "ontological" qualification of negative commons; he argues that such qualification depends on a preceding inquiry with the stakeholders directly concerned to determine whether objects hold a positive or negative value, which may lead to disagreements about the nature of the ruin itself, depending on the causes identified as responsible for its potential non-viability (Monnin, 2023c, p.50). The examples provided in this text are speculative and meant to make the categories more tangible to spark discussions in the context of the Limits workshop.

These questions are already emerging in the form of localized resistance movements contesting resource consumption of data centers in places like Latin America (Valdivia, 2024) and Ireland (Bresnihan & Brodie, 2025). Such political contestation is likely to spread, extending not only to the use-phase of ICT infrastructures but also to other resource-intensive stages of the supply chain, such as resource extraction and chip manufacturing, which rely heavily on water and energy (Crawford, 2019; Nova, 2022). Technologies like large language models or cryptocurrencies, due to their significant direct environmental footprints, may become focal points of these debates, as conflicting visions of their societal value emerge. For instance, while some digital currencies are embedded in proposals for universal basic income (Katz and Ferreira, 2020), others are viewed as emblematic of unsustainable trends of digitally-powered speculation.

Some technologies may also come to represent forms of material taboo in the context of a transition to sustainability oriented-futures: objects whose physical presence remains, but whose use, development or discussion becomes socially or politically unacceptable. As infrastructures associated with controversial or high-impact technologies, such as large-scale AI systems, are increasingly questioned on ethical or ecological grounds, they may be subject to processes of collective renunciation. In such case, the physical substrates of AI (servers, GPUs, training data centers and specialized hardware) could persist as inert remnants, even after the systems they supported are banned or widely rejected.

In his art piece *Newly forgotten technologies from AI-Free futures* (2024), critical artist and research Wesley Goatley imagines a future in which the term "Artificial Intelligence" itself has become taboo. He writes:

> "Ironically, in the years after the fall of A.I. it was rare to hear anyone mention those old A.I. voices at all. Names like Siri and Alexa became unheard of as baby names, being seen as bad luck, or in bad taste. This was part of a wider social taboo, an unspoken agreement that nobody would talk about the times when A.I. was always in the news and all over the internet, of what they had said back then, what they believed, what they bought, the warnings they ignored, and the things they allowed to happen. Just like the selective amnesia that obscures the trauma of childbirth, people both forgot and wanted to forget the time they all believed in A.I., and those imaginary friends. We had all moved on from that form of make-believe." (Goatley, 2024)

While Goatley's piece focuses on collective amnesia, it also helps think of the landscape in which the material elements that once supported AI remain visible yet holding negative connotations, as silent witnesses to a discontinued technological paradigm. These abandoned or repurposed devices, data centers, and support systems would constitute a subset of digital ruins: those whose materiality is not only obsolete, but stigmatized; culturally and politically marginalized due to their association with a discredited past.

### 3.1.2 Ghosts

As Monnin notes in his formulation of "ruinous ruins", material ruins do not exist in isolation: they are inhabited by the shadows of the systems that created them: expectations, knowledge, values and aspirations that may continue to shape social life and imaginaries long after the ruins have degraded or disappeared. These lingering elements, which I suggest to call *digital ghosts* here, form a second category of negative commons: cultural elements embedded in material histories. Digital ghosts will continue to exist in different forms, through memories, stories, documents, routines and institutional norms, as constant reminders of the possibility to engage with them in old or novel ways.

This category is especially important when considering the temporal lag between material disruption and cultural transformation. For instance, engineering knowledge needed to design high-performance chips, scale digital platforms, or manage global supply chains will not vanish simply because the infrastructures that relied on them are no longer sustainable. Similarly, workers whose identities and careers have been shaped by their contribution to digital systems may struggle with emotional and psychological dissonance in moving on from halted projects. As Goatley evocatively puts it, they may find themselves "living in the ruins of their old faith".

In that way, digital ghosts materialize a second set of pressures that is important to consider for societies in transition: even as infrastructures decay or are proactively dismantled, the cultural frameworks that shaped and legitimized them (e.g., ideas of growth, efficiency, innovation) may continue to exert influence. This can subtly reproduce the logics of the dominant paradigm, even within efforts to move away from it. Ghosts demand distinct forms of engagement that are political, cultural and epistemological. Some may remain visible and subject to contestation; such as growth-oriented business models, technosolutionist assumptions, consumerist logics or universalis ideals rooted in colonial worldviews. These ghosts are likely to generate political friction, as their continued presence in policies, platforms, or institutional design comes into tension with norms and values from the emergent paradigm (e.g., post-growth political reforms or a call for pluralistic values). Others, by contrast, may become taboo like their material counterparts: deeply embedded cultural residues whose open invocation becomes socially or politically unacceptable, yet which persist through practices and infrastructures. Surveillance, control-oriented governance and some framings of "freedom" may fall into this category; disavowed publicly, yet structurally or rhetorically maintained in more hidden or euphemized forms.

In this sense, the redirection of computing toward sustainable futures needs to involve disentangling from inherited assumptions and values, not just material elements. The narratives that justified digital expansion, such as promises of development, modernization or competitiveness, will outlive the infrastructures they supported, with long-lasting repercussions for local communities, laid-off workers and affected ecosystems. Addressing ghosts, therefore, is not a matter of optimization or change, but of confronting the cultural inertia and historical lock-ins that may anchor future trajectories in the past.

**3.1.3 Seeds**

Switching to the emergent paradigm of the Two Loop Model, I propose to look at what may constitute the *seeds* of sustainability-oriented digital futures. This category (and the next one) is necessarily more speculative than the previous two, as it engages with less clearly defined objects. Still, it is possible to point to existing infrastructures, practices, and systems that may offer concrete entry points for imagining sustainable transitions.

As contrast to ruins, seeds refer to the physical and infrastructural elements that already exist within the dominant paradigm but may be seen as "embryonic form[s] of an alternative mode of production in the making" (Caffentzis & Federici, 2014, p. 95). Examples include community-based infrastructures like repair cafés, localized mesh networks, or distributed manufacturing systems such as Fablabs. These elements may already prefigure alternative relationships to technology that emphasize maintenance, mutualization, and sufficiency rather than growth or efficiency.

Keeping a strictly material lens might be limiting, as ruins themselves can host seeds when reinterpreted through the lens of another paradigm. The repurposing of decommissioned ICT equipment, the use of abandoned industrial infrastructure for localized production, or the reinterpretation of discarded electronics through *zombie media* (Nova & Roussilhe, 2020) aesthetics are examples of how the same material conditions can take on different meanings. This reflects a broader potential reorientation of technological "innovation": away from extractive growth models and toward limits-informed or need-based practices.

Technological alternatives often emerge from contexts of scarcity or exclusion, developing in response to constrained resource or systemic marginalization. The field of *frugal innovation* (Radjou & Prabhu, 2015), for instance, explores how communities creatively meet essential needs such as healthcare, education, or food access under resource-limited conditions. These innovative responses treat constraint not as a limitation but as a creative driver (Bhatti & Ventresca, 2013). Examples include the Zeer pot, a passive, electricity-free refrigeration system, which uses evaporative cooling to preserve food, and the Jerry Valentin initiative in Côte d'Ivoire, which refurbishes discarded hardware into functioning computers inside recycled containers, running on free and open-source software.

Beyond need-driven initiatives, other practices serve as intentional alternatives. Tomlinson et al. (2013) describe these as "practices of destabilization": efforts deliberately designed to redirect sociotechnical trajectories (p. 11). One example is the concept and community of practice of *permacomputing* (Viznut, 2020), which follows design principles inspired by permaculture that promote modularity, low-energy computing, repairability and the reuse of salvaged components. Grounded in principles like "care for life" and "keep it small", it offers an early model of ICT designed for futures defined by material limits. Not just pragmatic solutions, but culturally and politically charged practices, grounded in alternative imaginaries of care, limitation, and interdependence.

Seeds also include a wide range of initiatives currently operating more broadly within the dominant paradigm, embodying values aligned with sustainability and sufficiency perspectives. These include digital commons (e.g., open hardware/software, wikis, Creative Commons platforms), formal and informal repair infrastructures, and mutualization spaces such as Fablabs and Makerspaces. To be clear, not all Fablabs and repair cafes embody values of care, justice and sustainability. Many of them reproduce the very logics of the dominant paradigm (Kohtala, 2016). But as we move towards sustainability-oriented futures, they represent structures and practices that can be used and potentially fertile ground for redirection trajectories.

While these seeds already operate in the present, their trajectory through the transition from one paradigm to the other may follow different paths. Some will likely remain relatively stable, continuing to meet local needs with limited external support, as is the case for informal repair economics that already function autonomously in different contexts. Others may gain increased recognition or infrastructural support, as sustainability and resilience become formal institutional goals. Repair hubs, digital commons and low-tech networks may grow in prominence as their relevance to limited resource access becomes strategic.

Others may go through forms of adaptation and reinterpretation, as existing practices are adapted to fit new constraints or infused with new meaning. For example, older electronic components and devices previously seen as e-waste could find new purposes under a scenario where production or imports of new electronic devices is greatly reduced. Their value (as whole or as part) could increase significantly to allow digital practices based on reconditioned, repurposed or reassembled equipment; drawing from hacker cultures or repair aesthetics, and contributing to new forms of adaptation and creativity. Similarly, network infrastructure, storage systems and digital devices are unlikely to disappear entirely, but their governance and design priorities may evolve. Instead of centralized, resource-intensive models, we may see more localized, modular and resilient forms.

In this sense, seeds are critical to the transition to the emergent paradigm: they are already here, but their visibility, relevance, and social meaning may shift considerably. Along with their embedded *visions*, they could become more central as broader societal reconfigurations gravitate towards sustainability.

**3.1.4 Visions**

Like their negative commons counterpart, seeds are closely tied to *visions*: the cultural elements that anticipate, justify and accompany their material emergence. These include aesthetics, values, goals, economic models, political orientations and imaginaries of the future. While ghosts represent cultural residues from the declining paradigm, visions orient and justify the formation of alternative futures. They emerge from within the dominant system, even if often marginalized or overlooked. Their transformative potential, like that of seeds, lies in their existence despite less-than-ideal conditions. They are not speculative ideals detached from reality, but rather promising possibilities; propositions already being discussed, experimented, sometimes already firmly in place, by communities and social movements.

Many seeds discussed earlier are animated by these emerging visions. For instance, permacomputing developed as a deliberate response to the harms of dominant ICT paradigms, much like permaculture came about as a necessary critique of industrial agriculture. It brings with it a broader vision than the tools and services it offers, one in which care, sufficiency, low energy and resource use and system resilience displace growth, optimization and performance as guiding ideals.

Even in necessity-driven practices, such as those labeled as frugal innovation, visions play an important role. While such practices often arise under conditions of scarcity, they are anchored nonetheless in values of local contextualization, appropriateness, re-use and sufficiency over accumulation, unnecessary add-ons and tech escalation (Basu et al., 2013). As some authors have highlighted (see for example Stöber et al. (2022)), external factors can make frugal innovation initiatives more or less compatible with sustainability frameworks and the relationship between frugality and sustainability must be proactively formed (as opposed to an inherently positive relationship). But the visions embodied by such projects seem to offer promising avenues on which to build alternative paradigms, a fertile ground for reinterpretation as sustainability becomes a more dominant societal concern.

The way we collectively engage with visions, just like with seeds, may evolve along different trajectories during a transition to sustainability-oriented futures. Some visions may persist, remain stable or even gain in importance due to their alignment with those of the emergent paradigms. If we follow the assumptions that in addition to transitioning to societies compatible with ecological limits, there will be an emphasis on issues of justice and fairness, visions linked to the ethics of care (Gilligan, 1993), communing, conviviality (Illich, 1973) or pluriversalism (Escobar, 2018) would likely become more central.

Other visions might be the heirs of would-have-been ghosts, that have been reinterpreted in the transition from one paradigm to another. For example, the notion of efficiency – central to the dominant paradigm in ICT but historically turning gains into increased resource consumption (Coroama & Mattern, 2019) - could be reinterpreted within a sufficiency framework: how can best optimize ICT uses and equipment with a guiding goal of reducing overall (and not relative) resource use and environmental harms? Similarly, how can we rethink notions of equitable access to meaningful technology outside of the realms of personal ownership and device accumulation?

## 4 A PRELIMINARY FRAMEWORK TO BE DISCUSSED AND USED

Now that we have a clearer understanding of what ICT ruins, ghosts, seeds and visions could look like in the transition to sustainability-oriented futures, in what ways can this framework be useful? In this section, I will focus on two potential directions for this work to help participate in the discussions around redirecting technology towards sustainability-oriented futures.

### 4.1 In the context of the Limits workshop

First, this paper was developed in the specific context of the Limits 2025 workshop. I have attempted to build on multiple strains of work touching with issues of sustainability and computing to provide a framework that gives a new way to think about this intersection. I have then formulated four categories and described the ways both physical and cultural elements of ICT might evolve and materialize in the redirection of computing towards sustainability-oriented futures.

To make these categories more tangible, I have provided speculative examples, based on existing initiatives and what I consider to be plausible evolution scenarios. However, as research on technology otherwise is gaining momentum, I realize that many different interpretations remain possible, particularly situated within broader debates on economic bifurcation. For example, the technosphere of a society remaining within capitalism—such as imagined by propositions from circular economy or green growth—would differ significantly from one aligned with degrowth or postcapitalism principles. Even among more closely aligned sustainability-oriented trajectories, the role and shape of technology can vary considerably. Moreover, what is ultimately understood as a ruin or a seed will differ greatly depending on local contexts and perceptions of value.

Still, I saw merit in mobilizing this framework within the Limits workshop to see *if* we can identify areas of consensus, dissensus and new relevant questions for such qualification, as well for future implications. With this purpose in mind, I summarized some of the earlier propositions within a table (Figure 2) as a heuristic tool to guide discussions at Limits 2025 and beyond.

|        | Made taboo | Made conflictual |        | Reinterpreted | Remains or reinforced |
|--------|------------|------------------|--------|---------------|-----------------------|
| Ruins  | • Certain digital devices and uses<br>• E-waste | • Non-repairable devices<br>• Resource-intensive technologies<br>• Data centers<br>• Manufacturing sites | Seeds | • Localized production<br>• Network infrastructure<br>• Data centers<br>• Older electronic components<br>• E-waste | • Digital commons<br>• Community hubs<br>• Decentralized storage solutions<br>• Repair infrastructure<br>• Modular electronics |
| Ghosts | • Surveillance<br>• Control<br>• Freedom | • Growth-oriented business models and value systems<br>• Techno-solutionism<br>• Universalism<br>• Consumerism | Visions | • Innovation and progress<br>• Efficiency<br>• Access | • Care<br>• Commoning<br>• Conviviality<br>• Sufficiency<br>• Pluriversalism |

Figure 2: Anticipated ICT ruins, ghosts, seeds, and visions in transitions toward sustainability-oriented futures.

*The categories are speculative and intended to provoke discussion; they can be challenged, revised, and adapted to different paradigms.*

During the workshop, two lines of reflection were suggested to engage with the framework, using Figure 2 as a starting proposition:

(1) Can we collectively identify criteria for what sustainability-oriented futures involve, from which a normative classification could be elaborated to distinguish ruins from seeds and ghosts from visions?

(2) What elements currently in the table would you change, remove, or place in another category? What would you add?

## 4.2 In practice

Second, I believe this framework can be useful outside of research, for teachers and practitioners involved at the intersection of technology and sustainability, or more generally for those interested in sustainability-oriented tech. I suggest three non-exhaustive applications below.

Like many foresight tools and frameworks, the one detailed here can be used in educational environments, whether in academia or professional contexts, to help cultivate futures literacy and critical thinking around sustainability transitions in tech. Frameworks like this one can help broaden perspectives of the future, helping move beyond linear thinking patterns and domain-specific biases. Distinguishing between physical and cultural elements can support students and practitioners in understanding technological choices as part of a broader socio-technical logics and historically and geographically influenced value systems. It can also help imagine more tangible alternatives to dominant models of technology, focusing on existing seeds and visions rather than traditional notions of innovation as disruption. This shift could help reorient attention to the conditions of emergence and support, rather than the product of innovation itself.

The framework can also serve as a diagnostic tool to help (re)assess operations, design choices, or equipment management and to support reflective practice. Using the four categories as analytical lens, practitioners can map and assess their current operations, considering what to preserve, what to question, what to let go of, and what to help bring forward. For individuals and teams engaged in sustainability transition efforts, it can be useful to investigate which elements they currently depend on that may not align with the futures they aim to build, whether due to associated harms or their incompatibility with projected environmental scenarios. Example questions might include: Do our business models assume continuous growth and resource extraction? What infrastructures are we dependent on? Could they exist in a world structured by just and sustainable paradigms? In contrast, paying closer attention to seeds raises questions about what existing alternatives could already be integrated into operations, such as switching to open-source software, partially or fully canceling big tech subscriptions, contributing to local communities and platforms or reinforcing links with local needs and networks. The last two categories—ghosts and visions—can be just as useful for enabling reflective practice, especially within teams committed to meaningful change. Exploring the ghosts that shape our choices, such as cultural residues of growth, innovation or user engagement, can reveal the assumptions embodied in design decisions, often implicitly and sometimes never challenged. Identifying these ghosts is key to mapping obstacles to change and can serve as a guide to what paths *not* to follow when redefining practices. For teams seeking to participate in broader socio-ecological reconfigurations, it seems particularly relevant to examine the implicit and explicit values embedded in their work: e.g., what future is being imagined? What are its associated ghosts and visions? To what extent do they open space for more sustainable futures?

Finally, the framework could also be used more actively as a strategic guide for practitioners involved in organizational transitions, to help orient design processes toward longer-term sustainability directions. In this sense, seeds and visions can be mobilized explicitly to anchor guiding principles, such as sufficiency, care, collective benefit, resource limits and their associated material dimensions, into product roadmaps, architectural decisions, and UX practices. Ruins and ghosts can be used to anticipate resistance and to proactively develop strategies for managing decline and addressing inherited legacies. The categories can support trade-off decisions, clarify values and guide reflection on current and future practices: are we maintaining ruins and ghosts on life-support or are we helping grow seeds and let new visions flourish?

## 5 CONCLUSION

Redirecting computing in a way that better accounts for planetary boundaries and natural resources requires addressing technology as embedded within broader socio-economic dynamics, shaped by physical and cultural realities that act as both barriers and springboards for imagining alternatives. Envisioning a shift from a dominant, unsustainable system to one rooted in ecological and social sustainability demands that we engage not just with technical redesign but with the socio-political transitions that accompany it. Transition studies models, such as the Berkana Institute's Two Loop Model, help us recognize that such transformations involve dynamics of both continuity and abandonment, and that both ascending and declining trajectories require care, maintenance, and intentionality.

To support this work, I have proposed a framework organized around four categories – ruins, ghosts, seeds and visions – as a heuristic for identifying how specific physical and cultural elements of ICT may either resist or support the transition to sustainability-oriented futures. This lens clarifies that the future of technology will be shaped as much by what we let go of as by what we choose to carry forward.

Central to this proposition is the notion of "negative commons" (Monnin, 2023b): persistent material (ruins) and cultural (ghosts) artefacts from the dominant system that no longer serve their original purpose but still demand political attention, care, or active renouncement. In contrast, seeds and visions, refer to existing practices embedded within the dominant paradigm that already embody elements of social and environmental sustainability. These physical and material elements are not utopian ideals but already-existing projects and infrastructures that need support and participation to build the foundations of a new paradigm.

Transitioning towards technology *afterwards* requires new political forms capable of managing not just growth and accumulation, but also maintenance, repair, refusal and decline. As challenging as it may be to build a new paradigm, it is equally crucial to learn how to live with and take responsibility for the legacies we inherit, while consciously nurturing the futures we want to cultivate.

This article has aimed to clarify these propositions with the help of illustrative examples, while recognized that what qualifies as a seed, vision, ghost or ruin will vary depending on political, economic and ecological contexts as well as different value systems and normative assessments. Future research is needed to apply and adapt this framework within a range of perspectives, including degrowth, postgrowth, democratic economic planning and postcapitalist imaginaries.

## ACKNOWLEDGMENTS
The author acknowledges that the typology was first initiated with Krystof Beaucaire and Simon Tremblay-Pepin through a previous collaboration. He would also like to thank the participants of the graduate seminar "Tech Otherwise: Designs for Just Sustainability", at the University of Toronto, for the insightful conversations and feedback that helped shape this work.